\documentclass[sigconf]{acmart}
\usepackage{amsmath}
\usepackage{verbatim}
\usepackage{caption} 
\usepackage{subcaption}
\usepackage[utf8]{inputenc}
\usepackage[ruled,vlined]{algorithm2e}

\newcommand\gbm[1]{\textcolor{black}{#1}}
\newcommand\jjd[1]{\textcolor{black}{#1}}
\newcommand\liwei[1]{\textcolor{black}{#1}}
\definecolor{bcolor1}{rgb}     {1.0,0.0,0.0}
\definecolor{darkgreen}{rgb}     {0.0,0.5,0.0}

\AtBeginDocument{%
  \providecommand\BibTeX{{%
    \normalfont B\kern-0.5em{\scshape i\kern-0.25em b}\kern-0.8em\TeX}}}

\copyrightyear{2022}
\acmYear{2022}
\setcopyright{acmcopyright}\acmConference[CHI '22]{CHI Conference on Human Factors in Computing Systems}{April 29-May 5, 2022}{New Orleans, LA, USA}
\acmBooktitle{CHI Conference on Human Factors in Computing Systems (CHI '22), April 29-May 5, 2022, New Orleans, LA, USA}
\acmPrice{15.00}
\acmDOI{10.1145/3491102.3501850}
\acmISBN{978-1-4503-9157-3/22/04}

\begin{document}
\newcommand{\mysystem}{Bayesian Compass~}
%

\title[Investigating Human-in-the-Loop Optimization for Designing Interaction Techniques]{Investigating Positive and Negative Qualities of Human-in-the-Loop Optimization for Designing Interaction Techniques}




\author{Liwei Chan}
\email{liweichan@nycu.edu.tw}
\affiliation{%
  \institution{\scalebox{.95}[1.0]{National Yang Ming Chiao Tung University}}
  \country{Taiwan}
}

\author{Yi-Chi Liao}
\email{yi-chi.liao@aalto.fi}
\affiliation{%
  \institution{Aalto University}
  \country{Finland}
}

\author{George B. Mo}
\email{gm621@cam.ac.uk}
\affiliation{%
  \institution{University of Cambridge}
  \country{United Kingdom}
}

\author{John J. Dudley}
\email{jjd50@cam.ac.uk}
\affiliation{%
  \institution{University of Cambridge}
  \country{United Kingdom}
}

\author{Chun-Lien Cheng}
\email{liencc.cs08@nycu.edu.tw}
\affiliation{%
  \institution{\scalebox{.95}[1.0]{National Yang Ming Chiao Tung University}}
  \country{Taiwan}
}

\author{Per Ola Kristensson}
\email{pok21@cam.ac.uk}
\affiliation{%
  \institution{University of Cambridge}
  \country{United Kingdom}
}

\author{Antti Oulasvirta}
\email{antti.oulasvirta@aalto.fi}
\affiliation{%
  \institution{Aalto University}
  \country{Finland}
}

\renewcommand{\shortauthors}{Anonymized, et al.}

\begin{abstract}
Designers reportedly struggle with design optimization tasks where they are asked to find a combination of design parameters that maximizes a given set of objectives. In HCI, design optimization problems are often exceedingly complex, involving multiple objectives and expensive empirical evaluations. 
Model-based computational design algorithms assist designers by generating design examples during design, however they assume a model of the interaction domain. 
Black box methods for assistance, on the other hand, can work with any design problem. However, virtually all empirical studies of this human-in-the-loop approach have been carried out by either researchers or end-users. 
The question stands out if such methods can help designers in realistic tasks. In this paper, we study Bayesian optimization as an algorithmic method to guide the design optimization process. It operates by proposing to a designer which design candidate to try next, given previous observations. We report observations from a comparative study with 40 novice designers who were tasked to optimize a complex 3D touch interaction technique. The optimizer helped designers explore larger proportions of the design space and arrive at a better solution, however they reported lower agency and expressiveness. Designers guided by an optimizer reported lower mental effort but also felt less creative and less in charge of the progress. We conclude that human-in-the-loop optimization can support novice designers in cases where agency is not critical. 
\end{abstract}

\ccsdesc[500]{Human-centered computing~Systems and tools for interaction design}

\keywords{Interface Design; Bayesian Optimization; Human-in-the-loop Optimization; Multi-objective Optimization; Haptics; Touch}

\maketitle

\section{Introduction} 

One central problem in design is that of finding a satisfactory operating point in a multidimensional design space, one that balances trade-offs between relevant design objectives (e.g., \cite{10.1145/2047196.2047276, dorst2004problem}). 
Such an operating point can be obtained using different strategies. 
A common strategy is relying on prior experience, intuition, and a bit of trial and error.
Under such a strategy, the designer explores the space by gradually searching for suitable parameter values and assessing the observed trade-offs between the objectives.
This approach can be effective when the design space is simple or familiar. However, as a method, it is not reliable. 
It is sensitive to the level of skill and prior-experience of the designer as well as the complexity of the design problem at hand.
Moreover, it scales poorly and offers no guarantees that all reasonable options have been considered.
An emerging alternative strategy which we study in this paper is to use an optimization-driven design method in which exploration is guided by a search algorithm \cite{bailly_menuoptimizer_2013,dudley_crowdsourcing_2019-1,guo_vinci:_2021,odonovan_designscape:_2015, oulasvirta2020combinatorial,todi_sketchplore:_2016}.
An optimization-driven design method guides the designer in their design space exploration and may offer various tools to inform final design selection.
In this paper we contrast these two approaches in an empirical study in order to report on the various positive and negative qualities of human-in-the-loop optimization.

Both the designer-led and optimization-driven strategies have conceivable advantages and disadvantages, thereby offering a rich collection of hypotheses worthy of examination.
With complete freedom over the exploration of the design space, the designer is likely to have a stronger sense of agency which may deliver greater engagement in the task.
A recent study of designers' expectations about data-driven design raised the loss of agency as a concern \cite{gorkovenko2020exploring}. 
On the other hand, a potential drawback of the designer-led approach is that exploration of the design space is either consciously or subconsciously constrained by preconceived notions held by the designer.
These preconceptions may be accurate, in which case constraints applied on exploration yield greater efficiency.
Empirical research has exposed biases that limit the creative capability, such as confirmation bias  \cite{hallihan2012confirmation}, 
as well as a tendency for \emph{fixation}, or `blind adherence with a solution' \cite{jansson1991design,youmans_design_2014}, which the literature suggests is hard to break \cite{agogue2014impact}. 
Promising regions of the design space can be missed and outcomes fail to deviate significantly from those arrived at early in the process by the designer.
We hypothesize that optimization-driven design may help to address problems such as design fixation but at the cost of designer agency and engagement.
Optimization-driven design also serves to mitigate sensitivity to the expertise and prior experience of the individual designer which in turn may deliver more consistent outcomes when engaging a group of designers of different skill and experience levels.

This paper contributes to empirical research on computational methods for designers. 
Our focus is on a HCI-related design task relevant for the development of interactive systems and interaction techniques.
Our anecdotal evidence is that relatively few papers presenting interactive systems at CHI, the premier venue of the HCI field, explore their parameter spaces systematically. 
We recognized the three following strategies described below.
First, potential design parameters can be assigned or eliminated by extrapolating from evidence presented in the literature.  
HiveFive \cite{HiveFive}, for example, is a VR visualization technique that was optimized by first referencing a biological theory of bee swarming to substantially narrow down the search range for each parameter, and second, fixing values in a pilot study (with three people).
Second, a divide and conquer strategy can be employed in which parameters are tackled one by one. 
For example, in Body Follows Eye \cite{BodyFollowsEye}, an interaction technique that guides users’ posture change in VR was optimized over a series of six sub-studies where each determined the threshold for one of the six motion types.
Third, sometimes the dimensionality of the problem is simplified with a mathematical model. 
For example, ErgonomicsTouch \cite{ErgonomicsTouch} exploits the so-called Hermite curve to amplify the user’s hand movements into a larger movement for increasing physical comfort while preserving ownership. 
It was optimized by reducing the dimensionality of the problem by identifying four parameters that determine the mapping curve, with respect to the objectives of accuracy, comfort, and ownership. 
Lower and upper bounds for amplifications were then determined empirically with a pilot study with five users.

To better understand the pitfalls and perks of optimization-driven design in contrast to a designer-led approach, we conducted a study with 40 novice designers.
We hypothesized that novices might benefit the most from computational assistance,
especially to achieve a degree of directedness and organization when exploring designs \cite{daly2012does}.
To account for learning effects across the two conditions, we used a between-subjects protocol assigning 20 participants to each condition and examined both the quality of design outcomes and the designers' subjective experience of designing.
The specific optimization technique we employed in the optimization-driven condition was Multi-Objective Bayesian Optimization (MOBO).
Bayesian optimization 
has shown significant potential in HCI design problems and offers an efficient method for exploring design spaces that are poorly understood by the designer at the outset.
To make this investigation concrete, designers are given a non-trivial design task involving the selection of parameters characterizing the behavior and haptic feedback of a 3D touch interaction in virtual reality to maximize efficiency and accuracy.
This design task involves two competing objectives for which the relationship to the controllable design parameters is unclear.
It therefore ensures a degree of challenge for designers and MOBO alike.

In summary, the core contribution in this paper is the empirical investigation of the positive and negative qualities of designer-led and optimization-driven design in a study with novice designers.
We found that the optimization-driven design of the 3D touch interaction technique delivers a superior outcome in terms of reducing spatial error but at the cost of the subjective experience of agency and ownership.
Furthermore, optimization-driven design using MOBO promotes wider exploration of the design space helping to mitigate detrimental design fixation.

\section{Related Work} 

Designing better interaction techniques is a long-standing topic within the HCI researcher and practitioner community.
This has motivated the development of various strategies and tools to support the designer in this process.
Papers in this vein in HCI typically demonstrate their new method or tool by highlighting improvements in the design outcomes but less commonly examine the secondary impact on the design process and the designer's experience.
In this paper we seek to understand how the interaction technique design process is influenced by the tools made available to the designer.
Specifically, we examine the advantages and disadvantages provided by human-in-the-loop optimization using Bayesian methods.

Below, we briefly review the related work to provide insight into the design process involving optimization methods within HCI.
We first cover the broader topic of data-driven optimization before examining interaction design with human-in-the-loop optimization and multi-objective optimization.
Finally we review prior work utilizing Bayesian optimization specifically to support the design process.

\subsection{Data-Driven Optimization}

One viable approach to improving interaction techniques is to leverage data collected on the whole or sub-tasks involved.
An example of this approach is provided by \citet{feit_toward_2017} who collected eye tracking data from 80 people performing a calibration task.
\citet{feit_toward_2017} demonstrated an optimization procedure leveraging this data to select optimal filter parameters and inform the design of gaze interfaces in terms of target sizes.

Captured data may also be combined with relatively simple empirical models such as Fitts' Law to optimize various interactive elements such as hierarchical menus~\citep{francis_designing_2000, matsui_genetic_2008} and keyboard layouts~\citep{ zhai_metropolis_2000, bi_quasi-qwerty_2010, dunlop_multidimensional_2012}.
SUPPLE~\citep{krzysztof_supple:_2004} takes a related approach in optimizing interface designs based on specified device constraints and user activity traces.
Deep neural networks modelling user performance when interacting with vertical menus~\cite{li_predicting_2018} have also been leveraged to drive optimization~\citep{duan_optimizing_2020}.
These various approaches may involve a degree of designer involvement to determine the feasible design space and interpret outputs, but the optimization process itself is largely offloaded to the computer.

Although not necessarily involving explicit optimization, data-driven methods leveraging deep learning have shown recent promise.
GUIGAN~\cite{zhao_guigan:_2021} employs a generative adversarial network (GAN) fed with a large dataset of real Android application graphical user interfaces (GUIs) to construct a generative model for creating novel application GUIs.
The quality of GUIGAN-generated GUIs is evaluated in the paper but there is no investigation of how the generative model can assist or influence the design process for designers.
Also employing deep learning, \citet{guo_vinci:_2021} introduce Vinci which applies a variational autoencoder to construct a generative model for advertising posters.
Critically, the Vinci system takes user input in the form of a product category, product image, and tagline text.
These inputs condition the generative process and are incorporated into the generated poster.
Various features of the Vinci system were evaluated with both novice and expert designers with generally favorable outcomes, particularly in terms of the tool's efficiency in generating a large number of design alternatives.
Nevertheless, concerns were raised by designers in terms of the ``controllability, comprehensibility, and predictability'' of the design process using Vinci.



\subsection{Human-in-the-Loop Optimization and Multi-objective Optimization}

Human-in-the-loop optimization refers to the process in which the optimization process is steered by human input, for instance through training feedback and observed human behavior to a set of input parameters.
This process has been extensively applied to HCI design tasks, for example in MenuOptimizer \citep{bailly_menuoptimizer_2013} where the designer is assisted during the task of combinatorial optimization of menus, and DesignScape \citep{odonovan_designscape:_2015} where layout suggestions for position, scale, and alignment of elements are interactively suggested to the designer.
Other design tools that have a human-in-the-loop aspect include Sketchplore \citep{todi_sketchplore:_2016} where real-time design optimization is integrated into a sketching tool; Forte \citep{chen_forte:_2018}, in which designers can directly iterate on fabrication shape design through topology optimization; in \citet{kapoor_interactive_2010}, where the behavior of classification systems can be iteratively refined by designers to support more intuitive behavior; \liwei{and in \citet{MultiArmedBandits}, where the arrangement of game elements is iteratively adjusted for increased user performance.} 
Overall, these tools all feature the central aspect of human interaction where the human actively participates during the optimization process to generate better designs. 
In broad terms, this human-in-the-loop paradigm of design is an evolution of the line of work introduced by \citep{myers_creating_1986} which aims to enhance the efficiency of the interface design process by automatically generating the code for the interface after demonstration of the interface specifications.

\jjd{\citet{yannakakis_player_2013} introduce the concept of player modeling in which a computational model is constructed of the cognitive, behavioral, and affective states of the player of a game.
This constructed model may be dynamically updated in-game based on observations of user inputs and, in turn, used to drive changes in gameplay and game content.}
\liwei{This general approach has been used to adjust game mechanics to maintain a challenging gaming experience for players~\cite{AdjustGame_2008, AdjustGame_2015}.}
\liwei{With a focus on designers as opposed to players, \citet{RolesOfAI_2019} explore co-creation with an agent for game level design and identify various potential roles for an agent in this design process, e.g., the agent portrayed as a friend, collaborator, student or manager.}
\jjd{\citet{liapis_mixed-initiative_2016} provide a review of related mixed-initiative methods applied to procedural content generation in game design.}

Multi-objective optimization for interaction design serves as a special case for optimization-based design where instead of one objective to optimize over, there are now multiple objectives.
As there is no longer one defined optimum for multiple objectives, the concept of Pareto optimality is important, where a design is considered to be Pareto optimal if no individual objective can be enhanced by changing the design parameters without resulting in at least one individual objective worse off. 
Multi-objective optimization aims to search for Pareto optimal designs so that an optimal trade-off between competing objectives is found.
In HCI, multi-objective optimization has been applied to touchscreen keyboard design to trade-off speed, familiarity, and improved spell checking \citep{dunlop_multidimensional_2012}, multi-finger input for mid-air text entry \citep{sridhar_investigating_2015}, and linkage design for a haptic interface \citep{hayward_design_1994}.
Many algorithms and computational methods have been applied for multi-objective optimization, including aggregating the different objectives into one via a linear weighted sum \citep{sridhar_investigating_2015}, grid-based methods \citep{dunlop_multidimensional_2012}, evolutionary-based methods \citep{knowles_parego_2006}, and Bayesian optimization \citep{hernandez-lobato_predictive_2016}.
In this paper, we seek to assess one specific multi-objective optimization algorithm, namely Bayesian optimization, in a human-in-the-loop context to explore the benefits and drawbacks as compared to the designer-led process, as it shows great potential in HCI design as detailed in Section \ref{ssec:related_work_bo}.





\subsection{Bayesian Optimization}
\label{ssec:related_work_bo}

Bayesian optimization is a machine learning technique for facilitating the optimization of unknown and/or difficult-to-evaluate functions.
It works by iteratively refining a surrogate model representing the function and intelligently selecting new test points to evaluate by balancing between exploration of the design space and exploitation of regions where the designs are particularly promising.
A major strength of Bayesian optimization is that the surrogate model is leveraged to ensure efficient search of the design space.
Bayesian optimization is therefore well suited to interaction technique design problems where the relationship between design parameters and user performance and/or subject experience is unknown or easily modeled.

Bayesian optimization has been employed in HCI to tackle various design problems as a human-in-the-loop optimization method.
Early work by \citet{brochu_bayesian_2010} demonstrated how Bayesian optimization can incorporate direct feedback from users in a preference gallery to help determine desired parameters governing the appearance of animations.
\citet{koyama_sequential_2017-1, koyama_sequential_2020} use a similar approach to allow users to rapidly adjust the visual appearance of an image in line with some desired aesthetic.
Bayesian optimization has also been used as a tool to determine game mechanic settings to maximize engagement~\citep{khajah_designing_2016}, adjust font parameters to maximize reading speed~\citep{kadner_adaptifont:_2021} and adjust interface and interaction features to minimize task completion time~\citep{dudley_crowdsourcing_2019-1}.
These various studies serve to highlight how Bayesian optimization provides an effective tool to support design tasks in HCI.
What is lacking, however, is a clear understanding of how design driven by this mechanism is experienced by or impacts the designer.



\subsection{Summary} 

The various research efforts reviewed above offer a range of alternative tools and techniques for optimizing user interfaces and interactions.
Lacking, however, is a clear understanding of how these various tools and strategies influence the design process and experience for designers.
This paper seeks to address this gap in the literature by comparing the outcomes and experience of designing with and without assistance from Bayesian optimization.
We focus on Bayesian optimization as the tool offered to designers given the significant advantages that have been demonstrated within the HCI domain in terms of its efficiency and its ability to handle black box optimization problems.

\section{Case: 3D touch interaction}
\label{ssec:3d_touch_interaction}
Our empirical study focuses on a complex and realistic interaction technique case -- 3D touch interaction -- which is ubiquitously applied in virtual reality.
Here, we compare two approaches: the designer-led and the optimizer-driven approach, and in this section, we outline the background of the interaction, the design space parameterization, and the design objective functions.
In particular, we specifically chose this task as 1) 
target acquisition in 3D is an important problem in the domain of virtual reality, 2) the resulting performance of the interaction is easily observable to the user as the design parameters vary, and 3) it serves as a classic multi-objective design problem in HCI as we will detail in Section \ref{ssec:background_3d_touch}.

\subsection{Background of 3D Touch Interaction}
\label{ssec:background_3d_touch}

\begin{figure*}[t!]
\centering
  \includegraphics[width=0.8\textwidth]{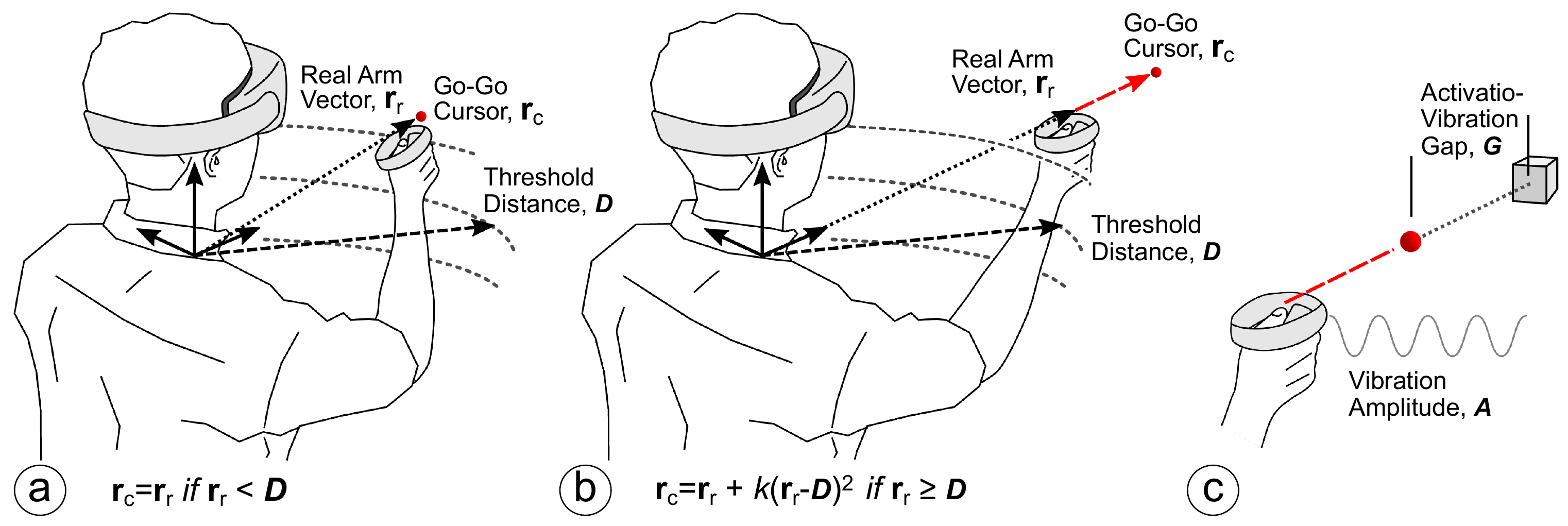}
  \caption{Our empirical study focuses on the task of improving the transfer function of the Go-Go technique. The technique calculates the virtual hand’s position with the parameters $D$ and $k$. (a) It maps the position linearly when the physical hand’s distance is within the range $D$, or (b) non-linearly by a factor controlled by $k$ when it moves beyond the range $D$. (c) In addition, the two parameters $G$ and $A$ for the activation-vibration gap and the vibration amplitude determine the vibrotactile feedback when the target is reached.} 
    \label{fig:gogo-design-params}
    \Description{The transfer function of the Go-Go technique calculates the virtual hand’s position with the parameters D and k. It maps the position linearly when the physical hand’s distance is within the range D, or (2) non-linearly when it moves beyond the range D. In addition, the two parameters G and A for the activation-vibration gap and the vibration amplitude determine the vibrotactile feedback when the target is reached.}
\end{figure*}

Target selection is a crucial, if not the most important, task for a virtual reality (VR) or an augmented reality (AR) application \cite{argelaguetsanz}.
A great variety of VR and AR selection methods have been proposed \cite{ 10.1145/323663.323667} in mind with the challenge of the trade-off between speed and accuracy that was identified in early works \cite{argelaguetsanz}.
Poupyrev categorized such selection techniques into the use of a virtual pointer or virtual hand metaphor \cite{POUPYREV199919}.
A good 3D selection design should allow selection to be fast and accurate; however, searching for the good design candidates while satisfying both objectives is known to be a challenging design problem.
Moreover, previous works showed that the control-to-display transfer function (including 2D and 3D selection) requires different numbers of parameters, which can range from two to ten \cite{argelaguetsanz, gogoTechnique, frees2007prism, meyer1988optimality, konig2009adaptive}. 
Thus, the high-dimensional design space makes searching a promising design instance especially time-consuming and costly.
For instance, previous approaches applied for designing 2D transfer functions are either based on a great amount of trial-and-error \cite{10.1145/2047196.2047276}, which is a costly process, or by heuristics \cite{10.1145/2766448, interactivity_crafter}, which requires prior domain expertise.

The \emph{Go-Go technique} is a well-known 3D touch interaction design proposed by Poupyrev, which has been widely applied in VR and AR interactions \cite{gogoTechnique}. 
Essentially, a hybrid control-to-display transfer function determines the virtual hand's position according to the physical hand's movement.
Within a certain range, the transfer function follows a linear mapping, in which the virtual hand moves linearly based on the physical hand's position.
Beyond this range, the transfer function follows a non-linear mapping, in which the virtual hand moves quadratically away according to the physical hand's position.
This combination enables users to stably touch the objects that are closer to the body while being able to hit the targets that are beyond the physical hand's reach.
Two parameters determine the switch of the mapping methods and the degree of the nonlinearity in the nonlinear schema.

Despite the number of the design parameters being relatively low, exhaustively searching the design space for the optimal design instance is not practical due to the challenges discussed above.
While the 3D touch interaction design is timely and increasingly important, optimization of its design either based on manual parameter tuning done by a designer or an optimization algorithm has not been well documented or explored.
For example, the Go-Go technique as described in the original paper recommends parameter settings without a proper rigorous justification \cite{gogoTechnique}. 
In the following experiment, we selected the Go-Go technique as a base example to compare a human designer's iterative search and the Bayesian optimization workflow, with some add-on design parameters to allow greater selection accuracy and speed. 

\subsection{Parameterizing 3D Touch Interaction and the Objective Functions}

\subsubsection{3D Touch Interaction}
The 3D touch interaction used in the later experiment is built on the original Go-Go technique with some modifications. 
The original Go-Go technique decided the chest position as the reference origin point. 
The arm vector $r_r$was obtained by subtracting the physical hand position to the chest then translating to the hand's coordinate and direction. 
In our experiment, we shifted the reference point to the shoulder, which captures more natural hand movements, as shown in Figure \ref{fig:gogo-design-params}. 
We further defined 1 unit of the ``operation range'' as the distance between the origin (which is shoulder of the operating hand) and the hand when the arm is fully extended. 
The Go-Go technique’s transfer function was then applied to calculate the virtual hand's position. 

\subsubsection{Design Parameters}
\label{ssec:design_params}
There are two parameters in the original Go-Go technique -- $D$ and $k$ -- which jointly form the hybrid transfer function. 
$D$ is the range which divides the linear and non-linear mapping, and $k$ determines the scale of the nonlinear component. 
If the physical hand's distance is within the range $D$, the transfer function linearly maps the user's physical hand to the virtual hand along the same direction, \liwei{where the real arm vector $r_r$ is assigned to the Go-Go cursor $r_c$ } (Figure \ref{fig:gogo-design-params}a).
Once the physical hand moves beyond the distance $D$, the nonlinear mapping allows the virtual hand to move much faster away from the origin (shoulder) along the direction of the physical hand by a factor controlled by $k$, \liwei{ with which the Go-Go cursor $r_c$ is computed as $r_r+k(r_r-D)^2$ (Figure  \ref{fig:gogo-design-params}b)}. 
We directly took $D$ and $k$ as the design parameters for our 3D touch interaction, and set the ranges of these two parameters to be $D \in [0, 1]$ and $k \in [0, 0.5]$. 


However, there are other parameters that will affect the 3D selection performance, including a vibration cue. 
This has been proven effective for enhancing efficiency and accuracy, and it has been applied to commercial devices.
Following this direction, we look to add the simplest and most pervasive haptic feedback when the target is reached to enhance user performance---a vibrotactile cue. 
For a balanced design, we selected two parameters for vibrotactile feedback: the activation-vibration point, $G$, and the vibration intensity, $A$, as shown in Figure \ref{fig:gogo-design-params}c. 
The duration of the feedback was fixed at 300 ms.
We set the range of the activation-vibration point to activate at any point in the range of 15 cm before and 5 cm after touching a target. 
We also set the vibration amplitude to be within the maximum voltage level (3.1V), which led the vibration amplitude to be within 2.6g. 
All design parameters are summarized in Table \ref{tab:3dtouch_parameters}.

\begin{table*}[t!]
    \small
    \caption{The four design parameters for the 3D touch interaction design, with the ranges. All four design parameters are continuous.}
    \vspace{-1em}
        \begin{tabular}{ l l c }
            \toprule
             \textbf{Design Parameter} &  \textbf{Description} & \textbf{Range} \\
             \midrule
             $x_1$: Distance Threshold, $D$ & Division between linear and non-linear mappings. & $[0, 1]$\\
             $x_2$: Scale Factor, $k$ & Scale of the non-linear component. & $[0, 0.5]$ \\
             $x_3$: Activation-Vibration Gap, $G$ & Cues when the target is reached. & [15\,cm, -5\,cm] \\
             $x_4$: Vibration Amplitude, $A$ & Vibrotactile feedback intensity. & [0\,g, 2.6\,g] \\
             \bottomrule
        \end{tabular}
    \label{tab:3dtouch_parameters}
    \Description{The table dictates four design parameters for the 3D touch design with the ranges. All four design parameters are continuous.}
\end{table*}

\subsubsection{Objective Functions}

The objective functions refer to the metrics we aim to maximize or minimize during the design process.
Following the discussion above, we considered two design metrics --- completion time (speed) and spatial error (accuracy) in target acquisition --- as our objective functions to be minimized.
The first objective function, completion time, refers to the average duration between the moment a target is shown in the 3D experimental environment and the moment it is successfully touched by the virtual hand. 
The second objective, spatial error, is the maximum overshoot distance, which is the maximum Euclidean distance between the virtual hand and the target’s 3D position if the virtual hand moves beyond the range of the target.
If a participant touches the target without any overshoot occurring (the cursor did not go beyond the range of the target at all), the spatial error will remain zero.

Because the values of the completion time and spatial error have their own ranges, normalization is required before the optimization process.
We converted these two metrics into two values which we refer to as \emph{speed} and \emph{accuracy} by linearly transforming the completion time ranged [1,600\,ms, 900\, ms] into to \emph{speed} ranged [-1, 1] , and the spatial error ranged [1\,cm, 0\,cm] into the \emph{accuracy} ranged [-1, 1]. 
Note that after the conversion, both the speed and accuracy objectives are now functions to be maximized instead (the higher value indicates better performance).
The ranges of completion time and spatial error were decided from a pilot test conducted with eight participants.

\subsubsection{Hyperparameter Setup for Bayesian Optimization}
\label{ssec:3d_touch_optimizer}

The Bayesian optimization in our implementation is built upon BoTorch\footnote{https://botorch.org/}, a PyTorch-enabled Bayesian Optimization library.
This library is commonly used in many research projects, and it offers reliable performance and the flexibility of picking the Gaussian Process models and acquisition functions.
The Gaussian Process we applied in the later experiment is the multi-output Gaussian Process. 
The acquisition function we applied is qEHVI, which represents the expected hypervolume increase, where we set $q=1$ to ensure that after each iteration, a batch of size one is selected to be given to the designer for testing.
Other hyperparameter settings include using 10 optimization restarts during the optimization of the acquisition function, 1024 as the number of restart candidates for the acquisition function optimization, and 512 as the number of Monte Carlo samples to approximate the acquisition function. 
These were selected to ensure good computational efficiency for each iteration of the optimization process.

\section{experimental method} 

The goal of the experiment is to investigate positive and negative aspects of human-in-the-loop optimization by contrasting it to the designer-led approach. 
The metrics we used to analyze the results cover the design outcomes and a wide range of designer experiences including the perceived creativity and workload. 
The optimization task consists of four design parameters left undetermined and two objectives to which the 3D touch interaction is set to be optimized during the design process.

In the designer-led condition, the search is progressed manually by actively exploring and refining design candidates. 
In contrast, the optimizer-driven condition follows a human-in-the-loop process in which a Bayesian optimizer leads the search for the designer; at the end, the designer determines the optimal designs from a set of Pareto optimal designs suggested by the optimizer. 
To avoid learning effects on the design target across experiment conditions, the experiment followed a between-subjects design. 
We measured the performance of designs produced in the two conditions, quantified the perceived creativity and workload using the Creativity Support Index \cite{csi-questionnaire} and NASA-TLX \cite{nasatlx-questionnaire}, and collected user feedback with a semi-structured interview. 
With the mixed-methods approach, we looked to understand the trade-offs for human-in-the-loop optimization as compared to the designer-led process.


\subsection{Participants}
We recruited 40 novice designers (20 F, 20 M), with a mean age of 22.2 years (sd: 2.4), \liwei{via snowball sampling and through a Facebook group page dedicated to recruiting participants from a local university}.
Most participants were enrolled in a master’s program with their expertise covering engineering, architecture, interaction, and education. 
Following the between-subjects design, they were randomly divided into the groups for the designer-led or optimizer-driven processes. 
All volunteered under informed consent and agreed to the recording and anonymized publication of results. 
They were compensated 20€ for their participation. 

\subsection{Apparatus}
The apparatus mainly consisted of the 3D touch interaction. 
However, the interface to support the optimization process was customized according to the experimental condition for the designer-led or optimizer-driven processes. 

\subsubsection{The 3D Touch Interaction and Prototype}
\begin{figure*}[t!]
\centering
  \includegraphics[width=1\textwidth]{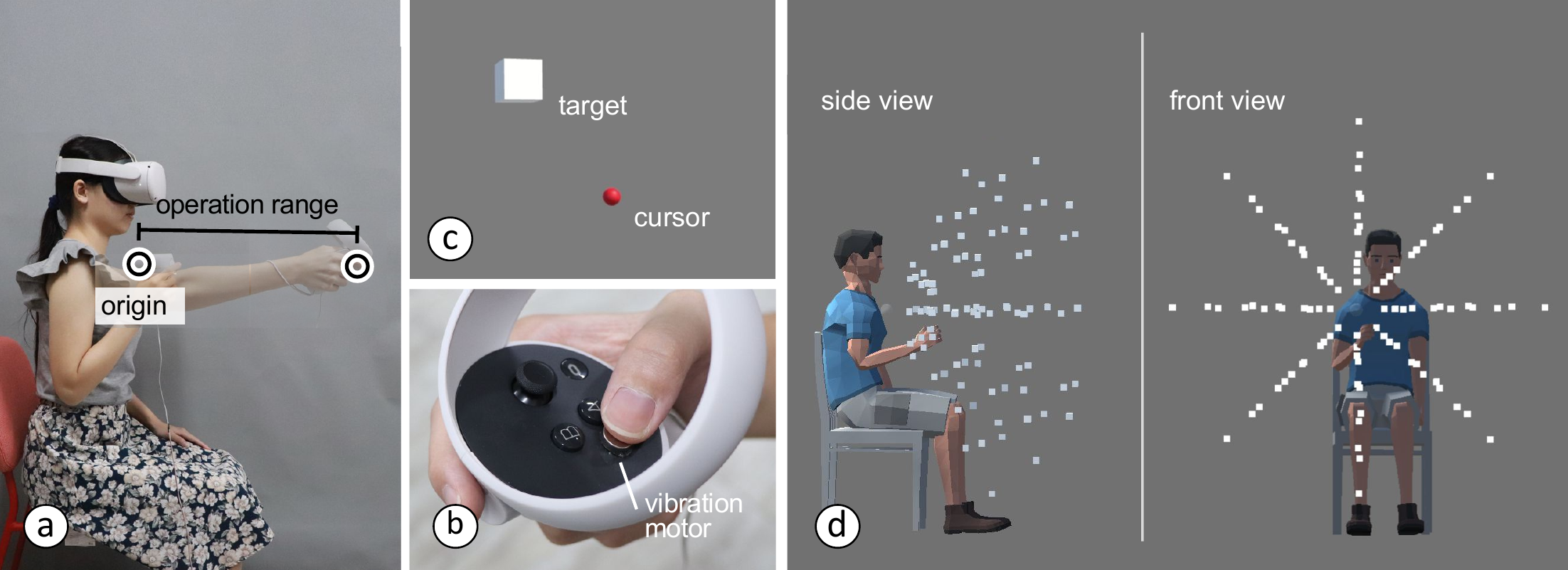}
  \caption{ (a) The experiment setup for the 3D touch interaction adapted from the original Go-Go technique, and (b) the interaction enhanced with vibrotactile feedback via the vibrator added to the controller. (c) Participants acquire the target using a cursor (e.g., the virtual hand) with dwell-based selection. (d) All possible locations of targets.} 
    \label{fig:3d-touch-setup}
    \Description{The experiment setup for the 3D touch interaction adapted from the original Go-Go technique and enhanced with controllable vibrotactile feedback via the vibrator added to the controller. Participants acquire the target using a cursor or the virtual hand with dwell-based selection. It also shows the 288 possible positions of the target.}
\end{figure*}

We built the 3D touch interaction in Unity 3D\footnote{\url{https://unity.com/}}  with the Oculus Quest 2\footnote{\url{https://www.oculus.com}} and the companion hand controllers, as shown in Figure \ref{fig:3d-touch-setup}. 
Our prototype implementation matches closely to the original one in \cite{POUPYREV199919} with the minor changes listed in Section \ref{ssec:design_params}.
To provide vibrotactile feedback on the controllers that can be precisely controlled, we added a vibration motor, Precision Microdrives 310-117\footnote{https://www.precisionmicrodrives.com/product/310-117-10mm-vibration-motor-3mm-type} (rise time 97 ms), on the controller such that users can easily rest their thumb on the motor. 
The vibrotactile feedback was controlled via a DRV2605L driver and an Arduino Uno microprocessor. 
During the optimization task, participants were asked to sit on a legged chair so that a polar coordinate system can be easily maintained. 
We followed task arrangements used in \cite{CHA2013350} for 3D target acquisition. 
The three variables that determined target locations are: the inclination angle (30\textdegree, 45\textdegree, and 60\textdegree); the azimuth angle (0\textdegree, 45\textdegree, 90\textdegree, 135\textdegree, 180\textdegree, 225\textdegree, 270\textdegree, and 315\textdegree); and the radial distance to the target (0.5 units, 1 unit, 1.5 units, 2 units of the operation range). 
The fourth variable determines target widths (3 cm, 4 cm, and 5 cm). 
In total, there were 288 (3 inclination angles $\times$ 8 azimuth angles $\times$ 4 distances $\times$ 3 target widths) variations of movement trials, as illustrated in Figure \ref{fig:3d-touch-setup}d. 

\subsubsection{The Parameter Sliders and Evaluation Button}
We offer parameter sliders and an evaluation button as shown in Figure \ref{fig:sliders-and-charts}a, which participants in the designer-led group use to adjust parameters for a new design and to initiate a formal evaluation of the design, respectively.
Four parameter sliders are located at the lower right-hand side of the participant in VR, whose values correspond to the four parameters of the interaction. 
Any adjustment of the slider values directly applies the design parameters to the interaction. 
Since there is always a random target presented in the virtual space, participants can test the current design by simply selecting the target; subsequently, the next target appears for further testing. 
To initiate a formal evaluation of the current design, the participant presses the evaluation button below the parameter sliders. 
This enters a dedicated mode where these widgets disappear and the participant starts to follow a series of 36 trials randomly selected from the 288 variations while keeping an equal sampling across target distance and target width. 
The evaluation was completed when 36 trials were finished. 
Then, the averaged completion time and spatial error of the trials were computed and indicated on the objectives chart (detailed in subsection \ref{ssec:parameter_chart}).

\subsubsection{The Parameters and Objectives Charts}
\label{ssec:parameter_chart}
\begin{figure*}[t]
\centering
  \includegraphics[width=1\textwidth]{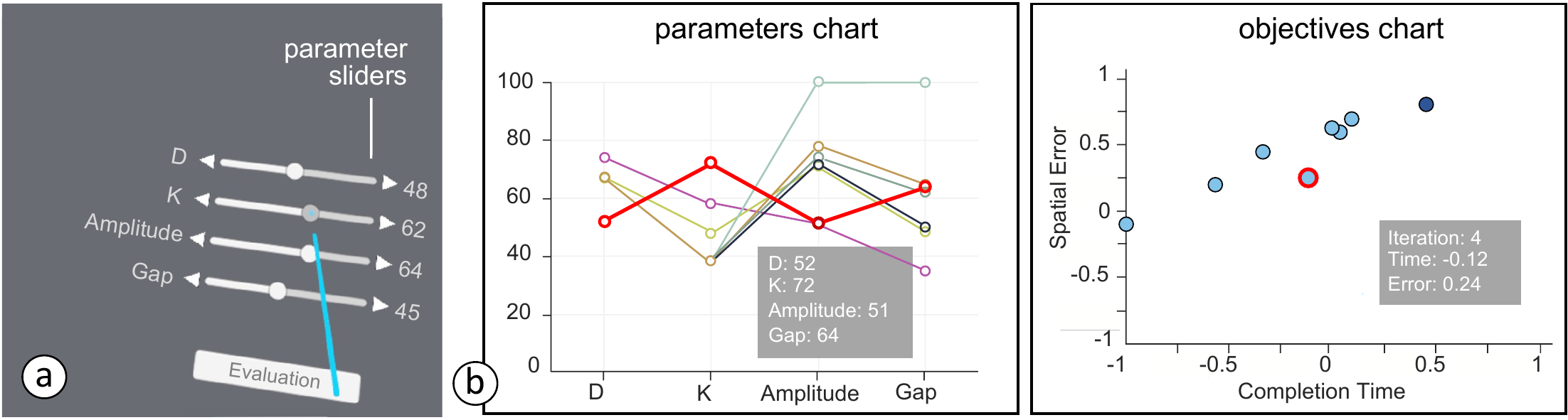}
  \caption{(a) In the designer-led condition, the designers can adjust the 3D touch interaction’s parameters using parameter sliders, and initiate a formal evaluation containing 36 trials on the current design with the evaluation button. (b) On completion of a formal evaluation, the parameters and objectives charts are brought up to show the evaluation results. The latest evaluation is indicated in dark blue, and the selected evaluation in red. 
} 
    \label{fig:sliders-and-charts}
    \Description{In the designer-led condition, the designers can adjust the 3D touch interaction's parameters using parameter sliders, and initiate a formal evaluation containing 36 trials on the current design with the evaluation button. On completion of a formal evaluation, the parameters and objectives charts are brought up to show the evaluation results.}
\end{figure*}

The parameters chart and objectives chart allow designers to keep track of all the designs that have gone through formal evaluation. 
The parameters chart contains a parallel coordinate plot of the designs evaluated, and the objectives chart contains 2D scatter plots of the corresponding objectives calculated from their formal evaluations. 
Once a formal evaluation is completed, the two charts are brought up for the participant to visualize the performance of the design under evaluation (Figure \ref{fig:sliders-and-charts}b).
The data point in dark blue in the objectives chart indicates the most recent evaluation.
Pressing the controller's menu button dismisses or invokes the charts.
These charts also support interactive functions. 
For example, the two charts are interlinked: on selection of a data point, indicated in red in the objectives chart, the corresponding design in the parameters chart is highlighted in red, and vice versa. 
Two floating text fields appear beside the selection to show detailed data of the evaluation.
In addition, the charts also directly apply the selected design to the parameter sliders and thus the interaction, allowing designers to easily revisit previously evaluated designs. 

\subsubsection{The Bayesian Optimizer}
In the optimizer-driven group, participants worked with the optimizer to determine optimal designs. 
The Bayesian optimizer was configured for optimizing the 3D touch interaction as described in Section~\ref{ssec:3d_touch_optimizer}. 

\subsection{Task}
We created a realistic brief for proposing 3D touch interaction designs in the form of a one-page description with background and goals.
Participants were prescribed as designers and were tasked to propose three optimal designs as the outcome of the design optimization. 

In the designer-led group, participants led the design process by actively testing and evaluating designs using the parameter sliders, evaluation button, and the charts. 
They were instructed to conclude the designs within a time limit of 60 minutes. 
However, they could propose to end early when they were satisfied with the design outcome. 

\begin{figure}[b]
\centering
  \includegraphics[width=1\columnwidth]{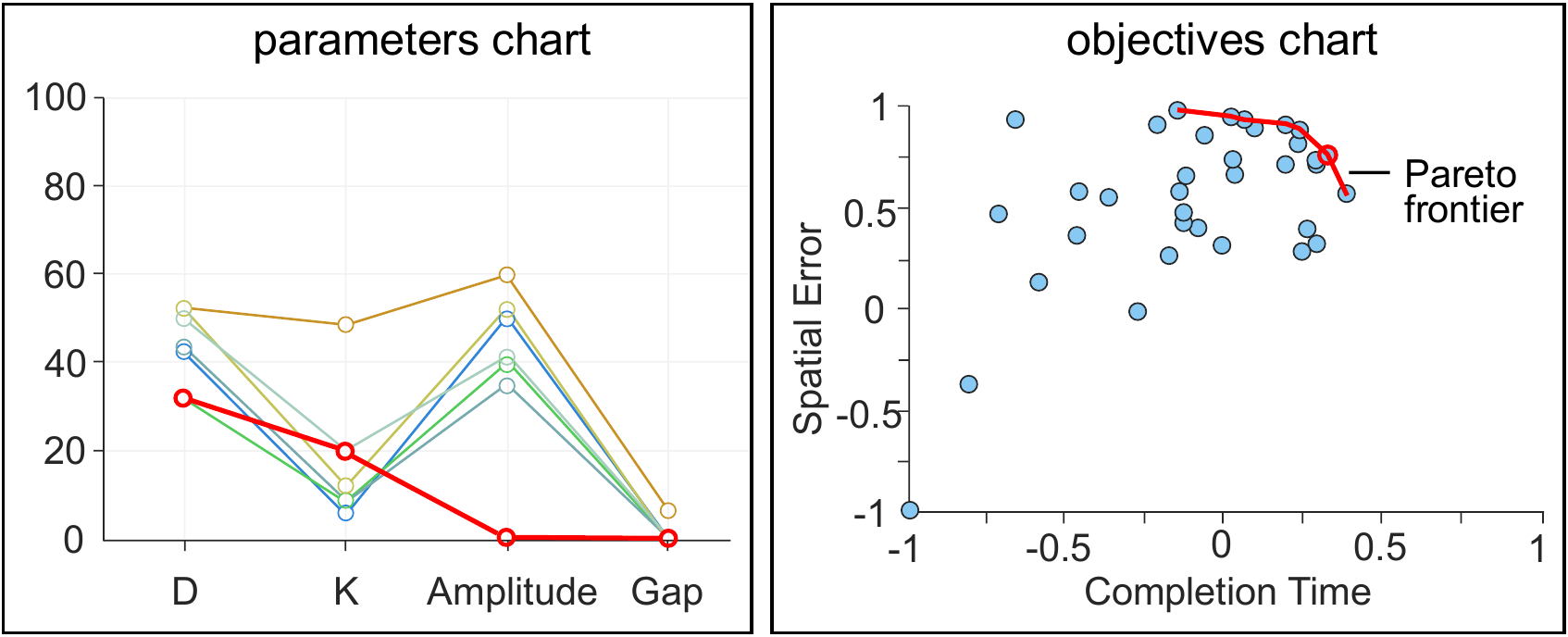}
  \caption{In the optimizer-driven condition, after the 40 formal evaluations, participants were allowed to test the Pareto optimal designs in the Pareto frontier, indicated in red.} 
    \label{fig:pareto-designs}
    \Description{In the optimizer-driven condition, after the 40 formal evaluations, participants were allowed to test the Pareto optimal designs in the Pareto frontier.}
\end{figure}

In the optimizer-driven group, participants worked with the optimizer in two stages -- the design and decision stages -- to conclude three optimal designs. 
In the design stage, the optimizer would propose in total forty designs; each required the participant to complete a formal evaluation by selecting 36 trials in sequence. 
After completing each evaluation, the design parameters and the design performance were displayed to the participant on the charts. 
The initial ten designs were randomly sampled by the Bayesian optimizer for optimization seeding. 
Completing the forty design evaluations entered the decision stage, where the participant was presented with the Pareto optimal designs (e.g., the designs connected by the red line on the objectives chart in Figure \ref{fig:pareto-designs}). 
They could test each of the Pareto optimal designs by selecting it. 
Then, they concluded the optimization process by selecting three designs from the Pareto optimal designs. 
As a result, the number of Pareto optimal designs could be fewer than three instances, in which case re-selection was allowed. 
In other words, if there was only one Pareto optimal design proposed, the three selected designs would be the same Pareto optimal design. 
From our study, the average number of Pareto optimal designs proposed is $3.3$ ($sd=1.5$) by the optimizer across participants.  

\subsection{Procedure}
\begin{figure}[b]
\centering
  \includegraphics[width=0.99\columnwidth]{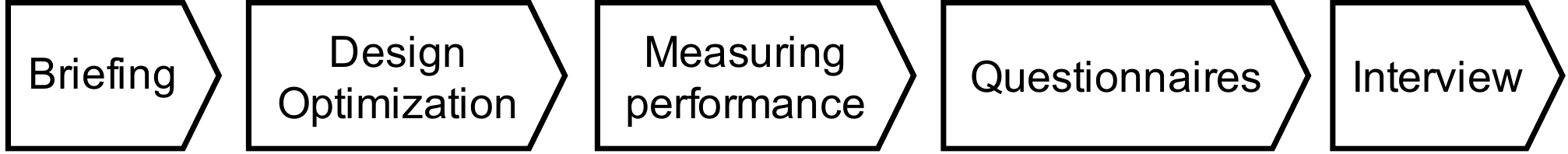}
  \caption{Diagram showing the study procedure: \textit{Briefing}; \textit{Design Optimization} where designers conclude three optimal designs; \textit{Measuring Performance} where design performance on the three designs is re-collected on designers; \textit{Questionnaires}; and \textit{Interview} } 
    \label{fig:study-procedure}
    \Description{The diagram shows the study procedure consists of five stages, which are briefing, design optimization, measuring Performance, questionnaires, and interview.}
\end{figure}

Figure \ref{fig:study-procedure} illustrates the study procedure. 
After briefing the study, the experimenter helped the participants wear the VR device, explained the parameters of the interaction, and allowed them to adjust the design parameters to observe the interaction behavior so as to familiarize the participants with the setup.
According to the participant's experimental condition, the experimenter introduced the interface and the overall procedure.
In design optimization, the designer-led group was tasked to propose three optimal designs within 60 minutes. 
The optimizer-driven group was told they would be working with an optimizer, which could take 60 minutes or longer depending on the situation. 

Once participants concluded their three designs, we again collected the performance data from them on those three designs in a separate session. 
Since the participants’ skill on the interaction may grow over time, this separate session was intended to ensure equal influence on the three designs’ evaluation. 
In this session, the three designs were presented in random order to the participant, each with a formal evaluation containing 36 trials to acquire their averaged performance.
Participants did not know which design among the three designs was under evaluation.

\subsubsection{Questionnaires}
We collected their subjective experience regarding the design process with three question sets. 
The overall experience set contained four 7-point Likert scale questions regarding  (1) Satisfaction: how much they were satisfied with the final design, (2) Confidence: how confident they felt the final designs proposed were optimal designs, (3) Agency: how much they felt they were conducting the design, and (4) Ownership: how much they felt they owned the final designs. 
We used the Creativity Support Index (CSI) \cite{csi-questionnaire}, a standardized psychometric tool for assessing the perceived creativity support of a tool.
It takes into account aspects of perceived creativity including exploration, expressiveness, results worth effort, enjoyment, immersion, and collaboration.
We also used NASA-TLX \cite{nasatlx-questionnaire}, a widely used assessment tool that rates the perceived workload of a task by looking at Mental Demand, Physical Demand, Temporal Demand, Performance, Effort, and Frustration.

\subsubsection{Semi-structured Interviews}
At the end of each experiment, we conducted a semi-structured interview focusing on experience, perceived issues, and how the participant values the design process and learns about the design space. 
The interview was audio-recorded. 
The procedure took about 2 hours in total per participant.


\section{Results} 

\subsection{Quantitative Results}

\subsubsection{Design Performance}

Figure \ref{fig:performance-and-experience}a shows the averaged completion time and spatial error of the three designs concluded by participant designers in each group. 
The average completion times were 1120 ms ($sd=119.4$) and 1185 ms ($sd=97.2$), and the averaged spatial errors were 2.2 cm ($sd=1.2$) and 1.5 cm ($sd=0.7$), in the designer-led and optimizer-driven groups, respectively. For statistical analysis, we initially log-transformed the completion time data, and confirmed the homogeneity of variances was not violated using \textit{Levene's Test} for both transformed completion time and spatial error data. 
Then, unpaired  t-tests were run on completion time and spatial error data to investigate if any significant differences exist between the groups. 
The analysis reported significant differences on spatial error ($t(38)= 2.237, p<0.05$) but not on completion time. 
This indicates the optimizer-driven method outperformed the designer-led approach in terms of the accuracy of the designs generated. 

\subsubsection{Designer Performance}

Notably, designers in the designer-led group spent 0.6 times less time in design optimization, but visited 6.7 times more design instances than those in the optimizer-driven group. 
The designer-led group participants spent on average 51.8 minutes ($sd=10.0$) on the design, compared to 78.0 minutes ($sd=6.3$) in the optimizer-driven group, \liwei{comprising on average 75.8 and 2.2 minutes respectively in the design and decision stages}. 

\subsubsection{Experience and Workload}

\begin{figure}[t]
\centering
  \includegraphics[width=1\columnwidth]{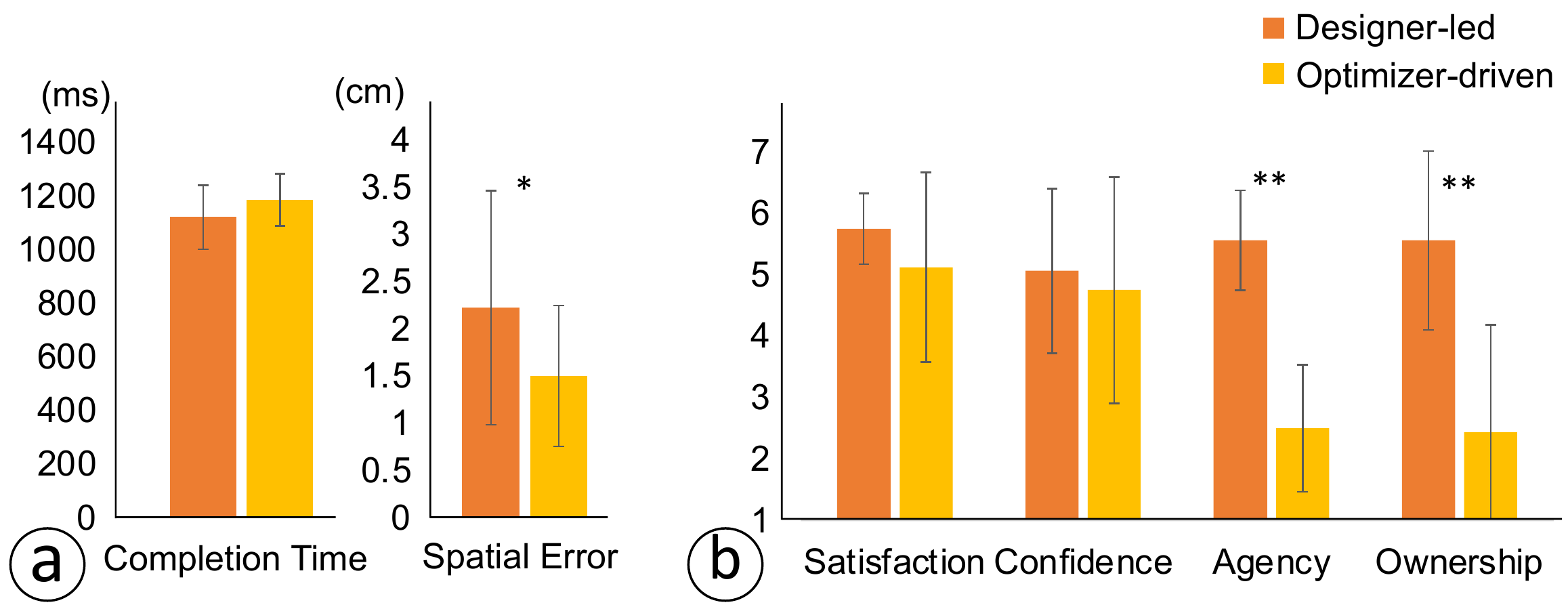}
  \caption{(a) The averaged completion time and spatial error of the designs concluded in the designer-led and optimizer-driven groups. (b) The ratings of general experience on Satisfaction, Confidence, Agency, and Ownership. The error bars denote 1 standard deviation. The one-star ($*$) and two-star ($**$) symbols indicate $p<0.05$ and $p<0.001$ significant differences, respectively.}
    \label{fig:performance-and-experience}
    \Description{The first chart displays the averaged completion time and spatial error of the designs concluded in the designer-led and optimizer-driven groups. Significant difference was found on spatial error only. The second chart shows the ratings of general experience on Satisfaction, Confidence, Agency, and Ownership. Significant differences were found on agency and ownership only.}
\end{figure}

Figure \ref{fig:performance-and-experience}b displays user ratings on Satisfaction, Confidence, Agency, and Ownership as well as the statistical analyses between the two groups.
We ran \textit{Mann-Whitney U Test} on each scale to investigate if significant differences exist. 
The analysis reported differences existed on Agency ($t(38)= -5.523, p<0.001$) and Ownership ($t(38)= -3.892, p<0.001$), but not on Satisfaction and Confidence. 

Table \ref{tab:csi-table} summarizes the CSI scores and the statistics analysis between the groups. 
The \textit{Mann-Whitney U Test} was applied on the overall CSI score and each factor comprising the CSI. 
The analysis shows a significant difference on the overall CSI score ($t(38)= -2.503, p<0.05$), suggesting that perceived creativity support was higher in the designer-led group than that in the optimizer-driven group. 
Comparing each of the factors, significant differences were only found on the Expressiveness factor ($t(38)= -3.222, p<0.001$). 
No differences were found on Exploration, Result Worth Effort, Immersion, and Collaboration.

The NASA-TLX scores and the statistical analysis between the groups are summarized in Table \ref{tab:nasatlx-table}. 
The \textit{Mann-Whitney U Test} was applied on the overall NASA-TLX score and each factor of the NASA-TLX. 
The analysis shows no difference in the overall score. 
Looking into each factor, significant differences were found only on the Mental Demand and Effort (both $p<0.05$). 
No differences were found for the Physical Demands, Temporal Demands, Performance, and Frustration. 
We found rationales that suggest the factor ratings in each group are distinct and worth discussion. 
In the following subsection, we will discuss the results and rationales between groups by factor. 

\begin{center}
\begin{table}[b]
\resizebox{1\columnwidth}{!}{\begin{tabular}{|l|l|l|l|l|l|}
\hline
                     & \multicolumn{2}{l|}{Designer-led} & \multicolumn{2}{l|}{Optimizer-driven} & Sig.          \\ \hline
Factor               & Score           & sd              & Score            & sd              & $p$            \\ \hline
Exploration          & 53.5            & 16.9           & 49.3             & 12.5           & .149          \\
\textbf{Expressiveness}       & 44.9            & 23.2           & 23.0            & 18.9           & \textbf{.001} \\
Worth Effort. & 55.7            & 22.9           & 48.6            & 26.2           & .301          \\
Enjoyment            & 44.0           & 28.1           & 40.8             & 35.6            & .678          \\
Immersion            & 21.4           & 21.0           & 28.2             & 18.5           & .183          \\
Collaboration        & 6.4             & 10.2           & 9.3              & 15.8           & .718          \\ \hline
\textbf{CSI}                  & 75.3           & 13.0           & 65.4            & 12.7           & \textbf{.011} \\ \hline
\end{tabular}}

\captionof{table}{User ratings on Creativity Support Index (CSI).}
\label{tab:csi-table}
\Description{The table lists user ratings of each factor in the Creativity Support Index.}
\end{table}
\end{center}

\begin{center}
\begin{table}[b]
\resizebox{0.95\columnwidth}{!}{\begin{tabular}{|l|l|l|l|l|l|}
\hline
                 & \multicolumn{2}{l|}{Designer-led} & \multicolumn{2}{l|}{Optimizer-driven} & Sig.          \\ \hline
Factor           & Score           & sd              & Score            & sd              & $p$           \\ \hline
\textbf{Mental.}   & 14.9            & 8.3            & 8.4             & 9.8            & \textbf{.011} \\
Physical. & 31.8           & 21.5           & 38.5            & 32.2           & .242          \\
Temporal. & 12.2           & 20.6           & 12.6             & 19.9           & .242          \\
Performance      & 25.1            & 19.8           & 15.7             & 12.7           & .398          \\
\textbf{Effort}           & 24.9           & 15.3           & 13.7             & 11.7           & \textbf{.040} \\
Frustration      & 8.5            & 12.7           & 10.0            & 15.8           & 1.00          \\ \hline
NASA-TLX         & 57.6            & 24.4           & 49.6            & 28.3           & .758          \\ \hline
\end{tabular}}
\captionof{table}{User ratings on workloads (NASA-TLX).}
\label{tab:nasatlx-table}
\Description{The table lists user ratings of each factor in the NASA-TLX.}
\end{table}
\end{center}




\begin{figure}[t]
\centering
  \includegraphics[width=1\columnwidth]{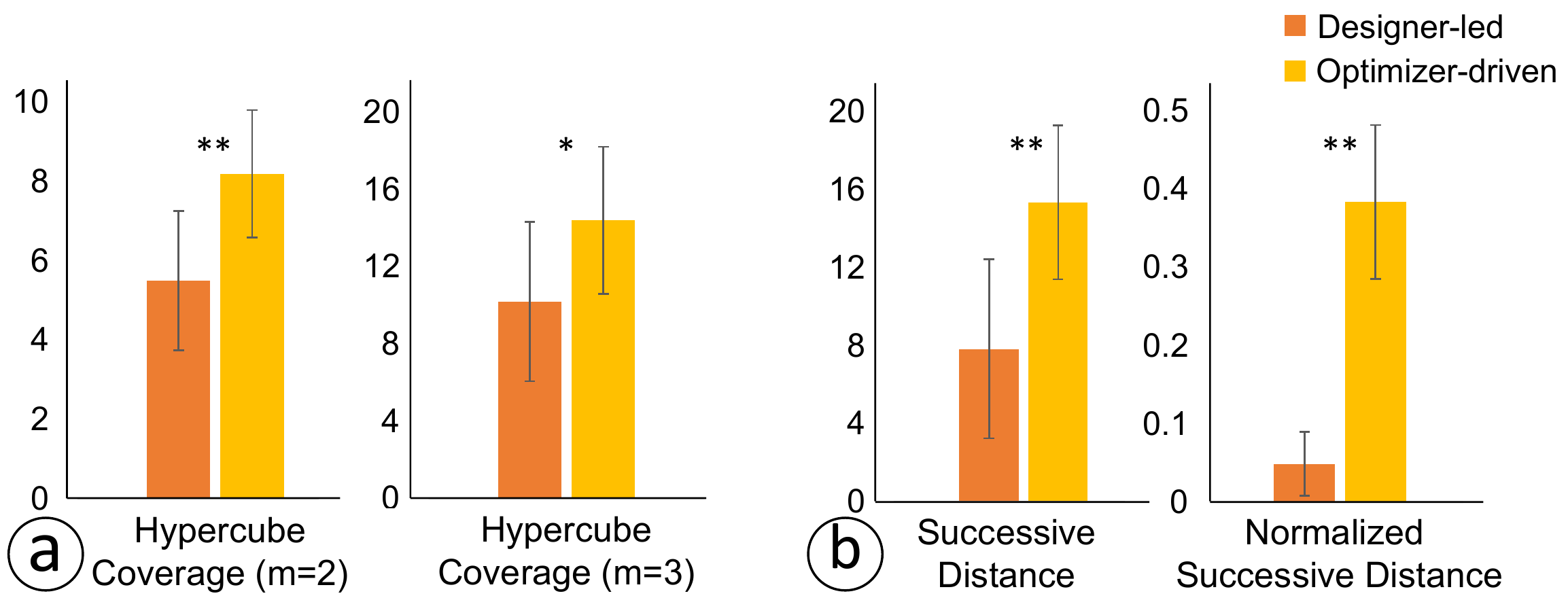}
  \caption{
  (a) The 
  number of hypercubes covered for both optimizer-driven and designer-led methods for $m=2$ and $m=3$. (b) The 
  total successive distance for both optimizer-driven and designer-led processes for the unnormalized case and the normalized case. The error bars denote 1 standard deviation. The one-star ($*$) and two-star ($**$) symbols indicate $p\leq0.001$ and $p\leq0.0001$ significant differences, respectively.
  }
    \label{fig:exploration-analysis}
    \Description{The first pair of figures shows the number of hypercubes covered for both optimizer-driven and designer-led methods for m = 2 and m = 3, both were found significantly different. The second pair of figures shows the total successive distance for both optimizer-driven and designer-led processes for the unnormalized case and the normalized case. Significant differences were found in both cases.}

\end{figure}





\subsubsection{Exploration and Exploitation during Design}
\begin{figure}[h]
\centering
  \includegraphics[width=1\columnwidth]{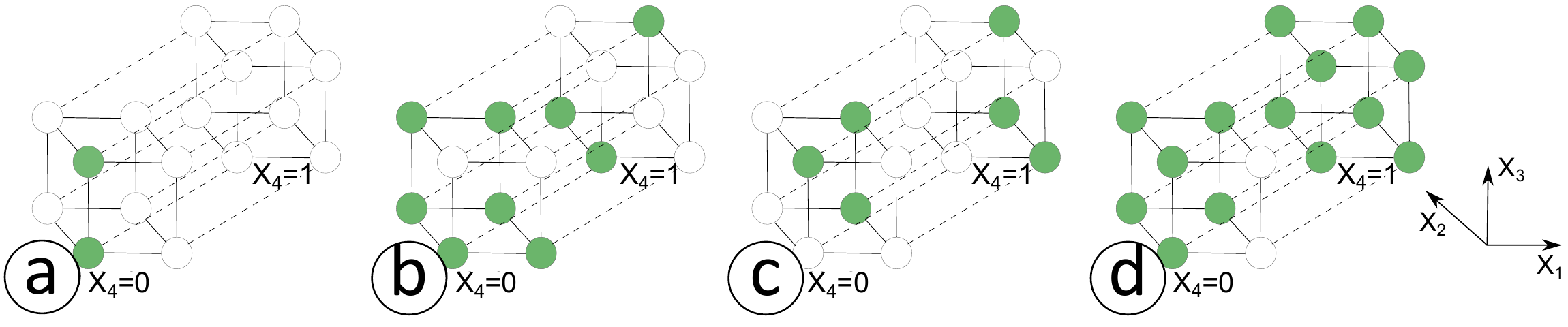}
  \caption{Figure showing the best and worst performance of hypercube coverage for both the designer-led and optimizer-driven conditions from the 40 participants. A hypercube is colored green if it is explored during the optimizer-driven or designer-led process. (a) and (b) show the worst and best coverage for the designer-led processes and (c) and (d) show the worst and best coverage for the optimizer-driven processes.}
    \label{fig:hypercube-compare}
    \Description{Figure showing the best and worst performance of hypercube coverage for both the designer-led and optimizer-driven conditions from the 40 participants. A hypercube is colored green if it is explored during the optimizer-driven process. They show the worst and best coverage for the designer-led processes and for the optimizer-driven processes.}
\end{figure}

In terms of design exploration, the designer-led group on average visited 271 different designs ($sd=192.4$), in which testing contributed on average 259 designs ($sd=194.5$) and formal evaluations contributed on average 12.5 designs ($sd=5.5$). 
In comparison, the optimizer-driven group visited only 40 designs selected by the optimizer. 
We further assessed how designers explored the design space in both conditions. 
To this end, we came up with the metric of finding how many hypercubes are covered.
For our specific application, the total design space is $[0, 1]^4$ and for a given division parameter $m$, we divide up the space into $m^4$ hypercubes.
We assign a hypercube as being covered if there exists a design parameter set obtained that lies within the hypercube bounds, lower bounds inclusive and upper bounds exclusive.
We have the upper bound being inclusive for the special case if the design parameter includes a parameter having the value of 1.
We assessed the hypercube coverage of both design methods with $m=2$ and $m=3$, each having 16 and 81 hypercubes respectively in Figure \ref{fig:exploration-analysis}a.
We see that for both values of $m$, the number of hypercubes covered is greater for the optimizer-driven process as compared to the designer-led method.
Figure \ref{fig:hypercube-compare} shows the hypercube coverage for the worst and best performances from the participants for both optimizer-driven and designer-led processes for $m=2$.
The figure illustrates that the worst-case and best-case coverage for the designer-led process covers less of the design space than that of the optimizer-driven process.
Furthermore, we conducted an unpaired t-test to assess whether the means of the two independent conditions are different, and we achieve a p-value of $0.0001$ for $m=2$ and $0.0019$ for $m=3$, both indicating very statistically significant results.
Therefore, this shows that optimizer-driven process is able to explore more of the design space consistently as opposed to the designer-led process and hence able to come up with more diverse design candidates.
This helps the designer in exploring more different candidates which can alleviate the problems of over-exploitation of a region in the design space. 

\gbm{
We also extended the hypercube coverage analysis for various levels of $m$ for the Pareto-optimal designs achieved by each participant.
For $m=2$, the mean hypercube coverage for the designer-led method is 1.4 (sd = 0.6) and for the optimizer-driven method is 1.5 (sd = 0.5).
There is no statistical significance in the difference of the means through an unpaired t-test through these two groups ($p=0.3950$).
For $m=3$, the mean hypercube coverage for the designer-led method is 1.7 (sd = 0.8) and for the optimizer-driven method is 2.0 (sd = 0.9), with no statistical significance in the difference of means ($p = 0.1766$).
However, as $m$ increases, the difference in the means becomes statistically significant as for $m=4$, the mean for the designer-led method is 1.7 (sd = 0.6) whereas it is 2.3 (sd = 0.9) for the optimizer-driven method with p-value of 0.0214, and for $m=5$, the means for the designer-led and optimizer-driven method are 1.7 (sd = 0.7) and 2.4 (sd = 1.2) respectively with a p-value of $0.0380$.
This shows the advantage of changing $m$ as a coarseness parameter in determining the level of exploration for different methods of interaction design, as $m$ increases, the hypercubes we considered to be covered become smaller in volume.}
The above analysis suggests that the optimizer-driven design may be better in determining a wider variety of Pareto-optimal designs with a statistically significant greater coverage of hypercubes as $m$ increases.
However, the region of the Pareto-optimal designs can also largely depend on the nature of the problem itself.
For instance in our application, certain parameters lead in general to better accuracy and speed trade-offs, and also variation between individual performances of different users.

Next, we assessed explicitly how much designers exploit narrow regions of the design space. 
We used the metric of the total successive distance---the sum of the Euclidean distances between successive design parameters tried for consecutive design iterations---to measure this.
If a designer is over-exploiting or fixated, the successive distance between designs would be small as opposed to a designer who is exploring many very different design candidates.
More specifically, a designer that would be fixated would focus on a smaller region of the design space, yielding design instances that are clustered to each other.
This results in a smaller successive distance between consecutive iterations and hence a smaller total successive distance.
If there was more exploration done on the design space, then the design instances would be in more disparate regions of the design space, yielding a greater successive distance between consecutive iterations and hence a greater total successive distance.
In addition, for the designer-led group, there are cases where the total number of design parameters attempted is very large (up to 830 iterations for both exploring and testing), whereas for the optimizer-driven group, the total number is set to be 40 iterations.
To account for the variation in the total number of design iterations, we also normalized the total successive distance over the total number of design iterations.
This metric would help eliminate the increase in the total successive distance due to simply more design iterations attempted.

The results for the successive distances are shown in Figure \ref{fig:exploration-analysis}b.
We see that for both normalized and unnormalized successive distances, the optimizer-driven process has a higher value than the designer-led method.
It is also worth noting that for the designer-led method, the variance in total unnormalized successive distance is similar to that of the optimizer-driven method, suggesting that both methods yield a similar level of exploration with respect to its mean successive distance.
Furthermore, we conducted an unpaired t-test to assess whether the means of the two independent conditions are different, and we achieve a p-value of $<0.0001$ for both the unnormalized and normalized total successive distances, both indicating statistical significance.
This shows that the optimizer-driven process leads to less fixation on a specific design region and with greater variation in terms of design exploration due to the greater discrepancy in the designs generated in consecutive iterations.
Therefore, this indicates that the optimizer-driven process is a useful tool for designers in order to cover more diverse design instances.




\subsection{Qualitative Results}

\subsubsection{Exploration}
18 out of 20 participants in the designer-led group stated the tool as intuitive, calling it “\emph{straightforward}” (P3, P7, P14)  and “\emph{easy to learn}” (P2, P5, P17).
Six designers stated the tool allowed them to be efficient at exploration (P3: “\emph{testing a design allowed me to gain some idea about the design before going into full evaluation}”) especially “\emph{when you want to quickly test alternatives around a design}” (P7). 
However, six participants reported some sort of anchoring bias, stating “\emph{I invested most of the time in fine-tuning.}” (P3, P5), and that they were aware that “\emph{many [alternatives] were left unvisited}”. 
P12 stated being stuck: “\emph{I think I can push further the [completion] time, but I can’t find how}”. 
P14 expressed dissatisfaction but was also resistant to re-initiate the search, saying “\emph{I may start over with any different design, but that would be another long investment}”. 
Designers in the optimizer-driven group perceived the exploration differently. 
Four participants stated it was interesting to watch “\emph{what designs the AI will bring up to me}” (P22, P24, P34, P38). 
P21 mentioned “\emph{it was obvious to me there were many different designs}” and stated he got to know the design space and established what constituted smooth interactions in the process.
These comments were echoed by P27 who commented: “\emph{experiencing bad and good designs is helpful in gauging how parameters gave good interaction.}”

\subsubsection{Explainability and Reflection}
Most participants stated the optimizer generally led them to better designs over time. 
However, there were those moments they would become confused when “\emph{the new  proposal suddenly appeared to be worse}” (P30). 
P24 mentioned “\emph{I thought I was doing good with the AI, but then it seemed to steer into a very different design direction}”. 
Some blamed the confusion on the AI side, thinking it was “\emph{broken}”, and “\emph{got lost}”. 
Others attributed the confusion to themselves, saying “\emph{I wondered if it was my [bad] performance that caused the AI to bring the design}” (P22). 
Ten participants stated they looked to have some form of explanation from the AI. 
In most cases, participants realized the optimizer steered them back on good-performance designs and could regain their satisfaction with the AI. 
Otherwise, two designers who ranked low satisfaction and confidence, commented “\emph{the AI was limited}” (P22, P34). 

\subsubsection{Agency and Expressiveness }
In contrast to the designer-led group, the designers in the optimizer-driven group generally expressed low agency and low expressiveness. 
Six designers stated they wanted to have some form of agency and express their ideas to the optimizer, especially when they disagreed with designs offered by it. 
For instance, P24 mentioned, “\emph{I knew what I wanted. I wanted the gap [value] to be reduced, but the AI didn’t give me that design}”. 
He suggested a feature of recommending the direction of adjustment, taking the gap as an example. 
Also, P32 suggested a feature for inputting preference on the design to AI, saying: “\emph{I wish I can just tell the AI I don’t like it [the design]}”. 
P33 wanted to skip evaluations where he thought  “\emph{trying out [in an evaluation] on a design that I knew wouldn’t work is a waste of time.}” 

\subsubsection{Ownership and Adaptability }
The optimizer-driven group received on average low ownership about the design outcome. 
However, participants reported mixed opinions, reflecting the relatively high variance in the ratings. 
Six participants attributed low ownership to low enjoyment, calling it “\emph{felt like working for the AI on those trials.}” (P22), “\emph{bored}”, and “\emph{not intellectual work}”. 
In addition, P28 commented on no sense of adaptation, stating “\emph{the outcome seemed not to reflect who I am}”, thinking others would also get the same design. 
Since the optimization algorithm leads the design, P30 stated, “\emph{the AI takes the responsibility of the design outcome}”. 
Four designers who gave high ratings commented about the concept of relatedness. 
For instance, P24 stated “\emph{I realized the AI was adapting design for me when I found the design is getting useful with increasing performance, that I felt I am part of the design}”. 
P35 mentioned the sense of relatedness saying “\emph{the AI was watching closely on those designs I performed well, and providing designs related, and it felt related to me}”.  
P38 attributed the ownership to the effort invested, “\emph{the AI cannot go on designing without me working out those trials}”.  

\subsubsection{Enjoyment and Engagement}
The rationales that suggest enjoyment are distinct between groups. 
In the designer-led group, participants enjoyed advancing the design outcome with their active exploration, saying “\emph{it resembles gaming}” (P12), and in particular, “\emph{seeing my adjustment result in progress is stimulating and helping me engage}” (P11). 
P4 said “\emph{although it's simple and repetitive, I don’t get bored on iterating.}” 
By contrast, designers in the optimizer-driven group attributed their enjoyment to curiosity and unexpectedness. 
Three participants stated, “\emph{an interesting way to learn design possibilities}” (P29) and, “\emph{fun to feel like working with the AI}” (P26). 
Four participants stated being suspicious, for instance saying “\emph{I was doubting it would work out}” (P22) but then felt excitement when seeing progress. 
Three said “\emph{you don't know what’s coming up next until you get to try it}” (P23, P28), so “\emph{each time got me something to expect}” (P38). P33 stated “\emph{adapting myself to a new design is the fun part and sometimes challenging}”. 
However, the enjoyment seemed to not last long; most participants mentioned the enjoyment reduced in later half rounds owing to long design time. 

\subsubsection{Effort and Responsibility}
The designer-led group perceived higher mental demands and effort invested than the optimizer-driven group. 
Four designers attributed the effort to “\emph{the need to figure out how each parameter works}” (P3), and “\emph{trying to further increase the performance}” (P14). 
Three participants stated it is challenging to handle two objectives, such that P18 commented “\emph{in fine-tuning, I tended to work on reducing completion time more than spatial errors.}” 
In the optimizer-driven group, participants reported mental effort was little. 
P22 stated “\emph{I feel relaxed as the AI is doing the design part}”. 
18 out of 20 participants ran overtime (more than 60 minutes). 
However, most reported little pressure of time. 
P24 mentioned “\emph{it was overtime but I didn’t feel it took that long}”. 
Two participants stated they did not feel responsible for the design outcome, saying “\emph{the AI took the lead and should take the responsibility}” (P32, P34).

\section{Discussion}
Our experimental results expose previously unreported trade-offs when using human-in-the-loop optimization to design interaction techniques. 
Differences found between the designer-led and optimizer-driven conditions are summarized in Table \ref{tab:takeaways}.
The results demonstrate that Bayesian optimization enables designers to explore the design space more broadly. In our study, optimizer-driven designers had around 1.5 times more extensive coverage when measured as hypercube coverage than when designers explore on their own. 
The optimizer-driven group also ended up with somewhat better designs. Their final designs better accounted for the balance of the two objectives with less effort, while designers without optimization assistance focused more on selection speed at the expense of accuracy.
However, on the negative side, optimizer-driven designers reported lower expressiveness and agency as well as lower ownership of the design outcomes. 
The low expressiveness and low agency are likely attributed to the fact that designers are `dictated to' by the optimizer resulting in a reduced sense of creativity. 
However, an observed benefit of this 'hand-holding' is that designers felt less effort: some attributed this to being more relaxed, while others felt less time pressure and less stress related to the design outcomes. 

\begin{table}[t!]
    \small
    \caption{Summary of differences found between designer-led and optimizer-driven conditions.}
    \vspace{-1em}
        \begin{tabular}{ c c c }
            \toprule
             \textbf{Factor} &  \textbf{Designer-led} & \textbf{Optimizer-driven} \\
             \midrule
             Completion Time & Equal & Equal\\
             Spatial Error & Worse & Better\\
            \midrule             
             Agency & Better & Worse\\
             Ownership & Better & Worse \\
             \midrule
             Exploration & Equal & Equal\\
             Expressiveness & Better & Worse \\
             Creativity Support & Better & Worse \\
             \midrule
             Mental Demands & Worse (High) & Better (Low)\\
             Effort & Worse (High) & Better (Low)\\
             \bottomrule
        \end{tabular}
    \label{tab:takeaways}
    \Description{The table summarizes the differences in qualities found between designer-led and optimizer-driven conditions.}
\end{table}

\subsection{Four Challenges to Improve Human-in-the-loop Optimization}

The results inform the development of better methods for human-in-the-loop optimization,
which in our view must converge proper interaction techniques with commensurate developments on the algorithmic side.

\paragraph{Challenge 1: Steering the optimizer with partial ideas.} 
Our results suggest that Bayesian optimization is effective when exploring a vast design space. 
A previous study on a system called Vinci, which used generative models to propose design suggestions interactively \citep{guo_vinci:_2021}, reported that designers felt a lack of diversity in important design dimensions.
However, our participants felt that loss of agency and expressiveness when being led by the Bayesian optimizer. 
We see this as an opportunity to develop interaction techniques that allow steering Bayesian optimization.

A key aspect of this challenge is to enable designers to express partial (vague) ideas that the optimizer could explore for them.
In our study, designers commented that once they had constructed an internal model of the requirements for a `good' interaction design, they wanted to be able to express these ideas to the optimizer. This was mostly strongly felt when they found themselves disagreeing with subsequent designs offered by optimizer. %
Reflecting on this feedback, interaction techniques are needed that allow users to express priorities in design dimensions, or directions where to look at next. 
However, such developments need commensurate developments in how the Bayesian optimization works, especially in the acquisition function.

\paragraph{Challenge 2: Mixed-initiative interaction.} 
Another direction to improve interactivity is to push the optimizer to the background, making its suggestions recommendations and not dictations as in our study. 
In a mixed-initiative fashion, it could make suggestions when it sees a significant opportunity.
For instance, the Bayesian optimizer could patiently construct a surrogate model of the design space in the background using only the evaluations the designers have encountered in the design process. 
If the optimizer observes that the designer is spending excess time examining a well-explored region of the design space, the optimizer can suggest alternative design candidates in less well-explored regions. 
This assistance could also be initiated by designers, for example by pressing a button to request a recommendation from the optimizer when they are stuck for ideas on how to improve the current design. 
Further, distinct support for exploitation and exploration could be offered for triggering recommendations that respectively aim for local improvements in regions of design space known to be promising or that aim to obtain new insight about unvisited or uncertain regions of the design space. 

\paragraph{Challenge 3: Improving transparency.}
Our designers expressed wanting the optimizer to be more transparent about the proposals. 
This finding is consistent with general observations within related research areas such as Interactive Machine Learning~\citep{dudley_review_2018}
 and Explainable AI~\citep{gunning_xai_2019}. User feedback indicated that designers expect monotonicity during the design process, meaning that designers expect that each new design proposed by the optimizer yields some improvement over the previous iteration. 
Confusion occurs when they experience the optimizer presenting designs that are then found to perform worse than preceding designs. 
This confusion in part stems from the users’ lack of knowledge about the inner workings of Bayesian optimization. 
It iteratively refines a surrogate model and leverages an acquisition function to drive the proposition of new points to test, in an exploration and exploitation trade-off. Exploitation seeks to sample where the surrogate model predicts a good objective while exploration samples where the uncertainty is high.
Transparency of the method could be improved simply by communicating in which mode it is currently operating so that designers then know they are assisting the optimizer in evaluating uncertain territory where high risk or opportunity is presumed.

\paragraph{Challenge 4: Supporting exploration/exploitation decisions.}
Our data suggests that user engagement comes from two sources: first, in exploitation where incremental improvement in performance can be expected, and second, in exploration where a fresh unfamiliar design attracts user attention. 
Human-in-the-loop optimization should help designers take these perspectives when needed. \liwei{A recent study \cite{InteractiveExploration} has explored this concept by allowing users to control sampling behavior in Bayesian optimization determined by acquisition functions so as to adjust the balance between exploration and exploitation}.
\liwei{Furthermore,} the participants commented that the exploitation process resembled computer games. In the optimizer-driven condition users linked unexpectedness to enjoyment. 
This observation suggests that it may be fruitful to encourage periodic switching between exploitation and exploration in order to improve engagement under both designer-led and optimization-driven strategies.
Such a control may be optionally applied to the Bayesian optimizer by simply assigning a minimum and maximum number of iterations spent in each of the exploitation or exploration modes before mode switching occurs.

\subsection{Limitations and Future Work}

Our findings are drawn from an empirical study on 3D touch interaction, of which the two objectives for optimization are clearly observable for human designers. 
Other types of interaction techniques that are not as perceivable to human designers may lead to different techniques to improve the optimization process, which calls for more experimentation.
In addition, the results of the empirical study are potentially subject to interpersonal differences due to the between-subjects protocol used. 
More experimentation is needed to validate reliability of the differences reported.

\section{Conclusion}
This paper has reported novel observations from a comparative study where two groups of novice designers, one optimized-led and the other self-led, completed a realistic interaction design optimization task.
Our main finding is that optimization-led design can help novices identify better designs, but at the expense of agency and expressiveness. 
When led by an optimizer, designers report lower mental effort but also feel less creative and less in charge of what happens. 
The results have a practical implication: designers who know a design domain poorly can benefit from Bayesian optimization when optimizing a design. 
However, more effort is needed to make optimization methods truly interactive, in particular in such ways that can help designers without compromising their agency over the process. 
We have proposed several ideas to this end in the previous discussion section.

\section{Open Science}
The Bayesian optimizer and the collected (anonymized) data are released on our project page: \url{https://userinterfaces.aalto.fi/dit}. Instructions for the prototype studied in the empirical part will be released, including the installation instructions and the computer program.

\begin{acks}
The research was supported by the Ministry of Science and Technology of Taiwan (MOST109-2628-E-009-010-MY3), Department of Communications and Networking (Aalto University), the Finnish Center for Artifcial Intelligence (FCAI), Academy of Finland (grants ‘OptiHAFE’ and ‘BAD’), and the Engineering and Physical Sciences Research Council (EPSRC EP/S027432/1). George B. Mo was additionally supported by a Trinity College Summer Studentship Fund.

\end{acks}

\balance
\bibliographystyle{ACM-Reference-Format}
\bibliography{acmart}


\end{document}